\documentclass[preprint,aps,amssymb,showpacs,superscriptaddress]{revtex4}
\usepackage{amsmath}

\newcommand{\lP}{\ell_{\mathrm P}}
\newcommand{\md}{{\mathrm d}}

\begin{document}

\preprint{AEI--2004--111}
\preprint{IGPG--04/11--7}
\preprint{gr--qc/0411101}

\title{ On Loop Quantum Gravity Phenomenology and the Issue of Lorentz Invariance}

\author{Martin Bojowald}
\email{mabo@aei.mpg.de}
\affiliation{Max-Planck-Institut f\"ur Gravitationsphysik,
Albert-Einstein-Institut,\\
Am M\"uhlenberg 1, D-14476 Golm, Germany}
\affiliation{Center for Gravitational Physics and Geometry,\\
The Pennsylvania State University,\\
104 Davey Lab, University Park, PA 16802, USA}

\author{Hugo A. Morales-T\'ecotl}
\email{hugo@xanum.uam.mx}
\affiliation{Departamento de F\'{\i}sica, Universidad Aut\'onoma
  Metropolitana Iztapalapa,\\ A.P. 55-534 M\'exico D.F. 09340, M\'exico}
\affiliation{Associate member of AS-ICTP Trieste, Italy}

\author{Hanno Sahlmann}
\email{hanno@gravity.psu.edu}
\affiliation{Center for Gravitational Physics and Geometry,\\
The Pennsylvania State University,\\
104 Davey Lab, University Park, PA 16802, USA}

\begin{abstract}
 A simple model is constructed which allows to compute modified
 dispersion relations with effects from loop quantum
 gravity. Different quantization choices can be realized and their
 effects on the order of corrections studied explicitly. A comparison
 with more involved semiclassical techniques shows that there is
 agreement even at a quantitative level.

 Furthermore, by contrasting Hamiltonian and Lagrangian
 descriptions we show that possible Lorentz symmetry violations may be
 blurred as an artifact of the approximation scheme.  Whether this is
 the case in a purely Hamiltonian analysis can be resolved by an
 improvement in the effective semiclassical analysis.
\end{abstract}


\maketitle

\section{Introduction}

One of the cases where observations of quantum gravity effects have
been imagined relies on modified dispersion relations for matter (such
as photons or neutrinos) travelling on a quantum gravitational
background \cite{AC1}. Planck scale effects are expected to be
negligible in standard circumstances, but observations of highly
energetic particles travelling long distances or other high precision
experiments may set bounds on possible effects.  Such effects have
been interpreted as indicating a violation of (standard) Lorentz
symmetry \cite{kostelecky1} for which, as expected, the observational
bounds are rather stringent \cite{bounds}. The theoretical foundation,
on the other hand, is still open and actively debated. For instance,
while trying to link the breaking of Lorentz symmetry to a privileged
frame has been argued to be in conflict with our current understanding
of field theory \cite{finetuning}, one can try to invoke a deformed,
rather than broken, symmetry in the form of Doubly Special Relativity
\cite{DSR}. Independently of these field theoretical
considerations, the task of candidate quantum theories of gravity is
to provide reliable estimates on the magnitude of expected
modifications to the standard dispersion relations to be compared with
observations.

One such candidate is loop quantum gravity
\cite{Loops} which leads to a discrete structure
of the geometry of space. This discreteness can be expected to lead
to small-scale corrections of dispersion relations, just as the
atomic structure of matter modifies continuum dispersion relations
once the wave length becomes comparable to the lattice size. There
have been several studies already which derive modified dispersion
relations motivated from particular properties of loop quantum
gravity \cite{GRB,Correct,QFTonCST}, but at this stage
the control on the calculations is insufficient. The difficulty lies
in the fact that loop quantum gravity is very successful in
providing a completely non-perturbative and background independent
quantization of general relativity which makes it harder to
re-introduce a background such as Minkowski space over which a
perturbation expansion could be performed. Techniques for
constructing semiclassical states are available and still being
developed further \cite{CohState,FockGrav}, but the calculations
toward modified dispersion relations are very complicated. Moreover,
the answer cannot be expected to be unique but to depend on several
parameters as well as quantization choices in the full quantum
theory.

In the first part of this paper we develop and study a simple model
which allows us to
introduce crucial properties of loop quantum gravity into the
Hamiltonian of a matter field.
As we will show, the model captures essentially all the effects
that have been considered
so far in loop motivated calculations of modified dispersion relations
even at a quantitative level. We can also see how different
quantizations would change the results, and which quantization choices
should have the largest effect on the order at which corrections
occur. Thus, we have the freedom to change basic objects according to
the possibilities of loop quantum gravity, but a much simpler and more
immediate way to check the consequences.

The essential idea in constructing the model is to consider space
as being made of homogeneous patches defining a lattice on which
the matter Hamiltonian, in particular its space derivatives, will
be discretized. This models the discrete structure of loop quantum
gravity, but could also be used classically as an approximation
(the metric field is then simply considered as a piecewise
constant rather than continuous function). Such a classical
approximation would become better and better if we choose smaller
and smaller patch sizes. A second ingredient from quantum geometry
then is that the patch geometry must be quantized (which is
readily done for homogeneous or even isotropic patches
\cite{Cosmo}). This implies additional, quantum
geometric corrections which grow with shrinking patches. Thus,
with effects from quantum geometry there is a non-zero patch size
leading to a minimal deviation of the effective matter Hamiltonian
from the classical one.

This model can be formulated at different levels of complexity
which allows to consider more realistic situations and also to
bring it closer to what one would have in full loop quantum
gravity. In this paper, we only consider the simplest construction
using isotropic patches of equal size, and only couple a scalar
matter field with simple discretizations. Still, as we will show
below, this simple model captures essentially all the effects that
have been considered so far in loop motivated calculations of
modified dispersion relations even at a quantitative level.
Although there are still gaps between the model and full loop
quantum gravity, it gives -- because of its simplicity -- a more
direct link between quantum gravity phenomenology and the full
theory.

In the second part of the paper, we study the possibility that the
apparent Lorentz violation of dispersion relations obtained in our
simple model, as well as in the more sophisticated treatments
\cite{GRB,Correct,QFTonCST}, be a result of the
approximation scheme rather than of the theory itself. Common to all
these computations is that they derive corrections to the matter
\textit{Hamiltonian}. This is natural in the setting of loop quantum
gravity because the quantization of spatial geometry is readily
available whereas 4d covariant quantities are harder to quantize. On
the other hand, by way of examples we will demonstrate that a purely
Hamiltonian analysis is much more subtle than a Lagrangian one, and
discuss how a perturbative Hamiltonian analysis can be improved in
order to draw reliable conclusions.

At first it might seem that Hamiltonian and Lagrangian
descriptions are completely equivalent, which certainly is the
case when theories are considered exactly.  However, as we will
demonstrate, the situation changes when approximation schemes, in
particular perturbation theory, are employed. When higher
derivative corrections are involved, the Legendre transform does
not commute with expansion in a perturbation series such that the
\textit{perturbed} Hamiltonian and Lagrangian formulations are not
necessarily equivalent. This is in particular the case for
theories non-local in time, in particular when time is not a
continuous parameter but `discrete'. This might well be the case
in quantum gravity and specifically in loop quantum gravity
\cite{cosmoIV}.  We will show in examples that going over to an
approximate continuum Lagrangian and then to the Hamiltonian
description will lead to a Hamiltonian that one could not have
obtained from a Hamiltonian model with continuous time in the
spirit of what has been proposed for loop quantum gravity.  Our
conclusion will therefore be that many of the calculations done up
to now (including the first part of this paper!) can only yield
preliminary results and that a definite answer to the question of
Lorentz violation by loop quantum gravity will have to await a
more complete treatment. We will conclude with an outlook on
possible strategies in this direction.

\section{Quantum Corrections to the scalar field Hamiltonian}

The simplest way to couple loop quantum gravity (LQG) to a free scalar
field is via its Hamiltonian
\begin{equation}
\label{hamiltonian}
 H= \int_\Sigma d^3x \left(\frac{1}{2}N((\det q)^{-1/2} p_{\phi}^2+ \sqrt{\det q}
q^{ab}\partial_a\phi\partial_b\phi)\right)
\end{equation}
where $\Sigma$ is a Cauchy surface of the space-time manifold $M$
(assumed to be globally hyperbolic).  The more complete treatments
in the literature proceed by quantizing the gravitational part of
this Hamiltonian with LQG methods \cite{QSDV} and then take
expectation values in a semiclassical state. That state comes with
a discretization of the spacial slice $\Sigma$ and as a
consequence, the partial derivatives in the classical expression
\eqref{hamiltonian} are changed to lattice derivatives.  Further
differences to the classical expression result from quantum
corrections to the classical values of $(\det q)^{\pm 1/2}$,
$q^{ab}$ upon taking expectation values of the corresponding
operators in the semiclassical state.

To compare the corrected Hamiltonian obtained by this method with
the standard expression \eqref{hamiltonian}, the lattice derivatives
in the former are expanded in a Taylor series to obtain, by
collecting the lower orders of this expansion and of the other
quantum corrections, an effective Hamiltonian. Plane waves are
solutions to the equations of motion generated by this effective
Hamiltonian and the corresponding dispersion relations contain
corrections compared to the standard one. We emphasize that in this
section we follow the standard procedure \cite{Correct, QFTonCST}.

To compute the corrections for the Hamiltonian in a simplified way,
let us now propose a model which includes the expected properties (and
which can always be made more complicated to be more realistic). Let
us discretize space into patches on which we can assume the geometry
to be approximately isotropic. Each patch $\alpha$ then carries two
real numbers, one for the scalar $\phi_{\alpha}$ and one for its
momentum $p_{\phi,\alpha}$, and an isotropic semiclassical quantum
state $\psi_{\alpha}$ for the geometry. (There is also a lapse
function, a real number, per patch which is not so important for our
purposes.)  This corresponds to a scalar on a classical geometry made
of patches of a size given by the expectation value of the volume
operator in the semiclassical state.

Again, corrections in the Hamiltonian are of two kinds: Since it contains
space derivatives of $\phi$, which have to be replaced by finite
differences, there is a discretization error. In addition, geometric
quantities like $\sqrt{\det q}$ and its inverse have to be replaced by
expectation values of the corresponding operators. Choosing each
patch to have coordinate volume one (otherwise, there would be
unnecessary coordinate factors), one obtains
\begin{equation}
\begin{split}
 H_{\rm disc}= \frac{1}{2}\sum_{\alpha} N_{\alpha} \Big[& p_{\phi,\alpha}^2
((\det q)^{-1/2})_{\psi_{\alpha}}\\
&\qquad + (E^I_iE^J_i/\sqrt{\det
q})_{\psi_{\alpha}}
(\phi_{\alpha+e_I}-\phi_{\alpha-e_I})
(\phi_{\alpha+e_J}-\phi_{\alpha-e_J})/4\Big]\,.
\end{split}
\end{equation}
Here, a subscript $\psi_{\alpha}$ means taking the expectation value
in the state $\psi_{\alpha}$, and $\alpha+e_I$ denotes the neighboring
patch in direction $I$. For isotropic patches, the expression
simplifies to
\begin{equation}
\label{ham2}
 H_{\rm disc, iso}= \frac{1}{2}\sum_{\alpha} N_{\alpha}\left[
 (p_{\phi,\alpha}^2 (p^{-3/2})_{\psi_{\alpha}}+ \frac{1}{4}
 (\sqrt{p})_{\psi_{\alpha}}
\sum_I(\phi_{\alpha+e_I}-\phi_{\alpha-e_I})^2\right]\,.
\end{equation}
where $p=a^2$ is the isotropic densitized triad component.

So far, there are many parameters to specify the background for the
scalar: For each patch we have a state, which is characterized by its
expectation value for $p$ and its spread. At first, one can assume
that all patches have the same values, which still leaves us with two
scales in addition to the Planck length and a wave length. Since the
difference corrections increase with the size of the patches while the
corrections for inverse powers increase with decreasing size, the first
scale, the scale factor $a$ of the isotropic patches, can be fixed by
requiring a minimal sum of those corrections.

Specifically, we can relate the discrete scalar field values
$\phi_\alpha$ to a continuous field $\phi(x)$ by
$\phi_\alpha=\phi(x_\alpha)$ where $x_\alpha$ is a point in the center
of patch $\alpha$. Expanding $\phi(x)$ we get
\begin{equation}
\label{taylor}
\begin{split}
\phi_{\alpha+e_I}=\phi(x_{\alpha+e_I})=\phi(x_\alpha)
&+(x_{\alpha+e_I}-x_\alpha)^a\partial_a\phi(x_\alpha)\\
&+\frac{1}{2}(x_{\alpha+e_I}-x_\alpha)^a
(x_{\alpha+e_I}-x_\alpha)^b\partial_a\partial_b\phi(x_\alpha) +\ldots
\end{split}
\end{equation}
Now let us assume that the $x_\alpha$ are the vertices of a regular
cubic lattice aligned with the coordinates on $\Sigma$. Then
\eqref{taylor} simplifies to
\begin{equation}
\phi_{\alpha+e_I}=\phi(x_\alpha)
+\partial_I\phi(x_\alpha)
+\frac{1}{2}\partial_I^2\phi(x_\alpha)
+\ldots.
\end{equation}
The squared differences in the Hamiltonian \eqref{ham2} are thus
approximated by
\begin{equation}
\label{laplace}
\frac{1}{4}\sum_I(\phi_{\alpha+e_I}-\phi_{\alpha-e_I})^2
=\sum_I\Big[(\partial_I\phi)^2
+\frac{1}{3}\partial_I\phi\partial_I^3\phi
+\ldots\Big](x_\alpha)\, ,
\end{equation}
i.e. we get the second derivative term we need plus higher derivative
corrections. Let us write
\begin{equation}
B:=\frac{1}{3}\sum_I\partial_I\phi\partial_I^3\phi
\end{equation}
for the first order correction. It is certainly not rotation
invariant, the symmetry having been broken by the introduction of
the regular lattice of patches. In a more realistic calculation
one would work with a random lattice or average over regular
lattices with different orientations to define the semiclassical
state. As we are only interested in an order of magnitude
calculation, we disregard this issue here.

For the corrections of the momentum term in (\ref{ham2}) we can use
earlier calculations for the inverse scale factor resulting in
$(p^{-3/2})_{\psi_{\alpha}}= a^{-3}+\Delta a^{-3}$ with a quantum
correction $\Delta a^{-3}$ which for larger $a$ is perturbative in the
Planck length. In a triad eigenstate \cite{InvScale}, those
corrections would be $p_{\phi}^2\Delta a^{-3}=cp_{\phi}^2 a^{-3} \cdot
\lP^4/a^4= c a^{-1}N^{-2}\dot{\phi}^2\lP^4$ (since
$p_{\phi}=Na^3\dot{\phi}$) with some constant $c$ which can be
computed once we make a choice on the explicit quantization, while a
coherent state \cite{Bohr} would result in $d^2\lP^2/a^4$
instead of $\lP^4/a^4$. (Thus, the Planck length would be
replaced by its geometric mean with the spreading scale $d$ of the
coherent state.)  Using a coherent state thus makes the correction
smaller, as expected, but we cannot yet tell the order in $\lP$ since
$a$ is not fixed. To do this, we minimize the total correction to the
Hamiltonian (\ref{ham2})
\begin{equation}
 c a^{-1}N^{-2}\dot{\phi}^2\lP^4+ aB
\end{equation}
with respect to $a$. This gives
\begin{equation}
 a=\left(\frac{c}{2}\lP^4
\frac{N^{-2}\dot{\phi}^2}{B}\right)^{1/2}
\end{equation}
for a triad eigenstate and
\begin{equation}
 a=\left(\frac{c}{2}d^2\lP^2
\frac{N^{-2}\dot{\phi}^2}{B}\right)^{1/2}
\end{equation}
for a coherent state.

A classical wave solution with our notation (e.g.\ dimensionless
coordinates) has the form
\begin{equation}
 \phi=\exp(i(ak\cdot x+N\omega t))
\end{equation}
if all patches have the same size $a$ (otherwise, we would have to sum
to get the physical distance in the argument).
Thus, we have an implicit expression for $a$
\begin{equation}
 a=\sqrt{(c/2)\lP^4 \omega^2\lambda^4/(16\pi^4 c' a^4)}
\end{equation}
where we have expressed $B$ as $c' (2\pi a/\lambda)^4\phi^2$ and $c'$ is
between 1/9 and 1/3, depending on the direction of propagation (this
combines the factor $1/3$ in $B$ with another factor of
$|k|^{-4}\sum_Ik_I^4$ which can be seen to be bound by $1/3 |k|^4\leq
\sum_Ik_I^4\leq |k|^4$). Thus we get
\begin{equation}
 a=\left((c/(32\pi^4c')\lP^4\omega^2\lambda^4\right)^{1/6}\approx
 \left((c/(8\pi^2 c'))\lP^4\lambda^2\right)^{1/6}
\end{equation}
where we used $\omega\approx2\pi\lambda^{-1}$ which is only approximately
true since we expect corrections to the dispersion relation (but
corrections here would only affect higher order terms).

Thus, the patch size is a weighted mean of the Planck length and the
wave length, with a large weight on the Planck length. This means that
we need rather small patches, and it strongly reduces the order of
expected corrections: From the inverse powers of $p$ we expect
corrections of the order
\begin{equation}
\label{fluctu}
\lP^4/a^4\propto (\lP/\lambda)^{4/3}
\quad\text{(triad eigenstate)}\qquad
\propto (\lP/\sqrt{d\lambda})^{8/3}
\quad\text{(coh.\ state)}\,.
\end{equation}
The order of the corrections from the $\sqrt{p}$-term on the other hand is
\begin{equation}
\label{lattice}
a^2/\lambda^2\propto (\lP/\lambda)^{4/3}
\quad\text{(triad eigenstate)}\qquad
\propto(\sqrt{\lP d}/\lambda)^{4/3}
\quad\text{(coh.\ state)}\,.
\end{equation}
These results square very nicely with \cite{QFTonCST}:
First of all in both cases there is a spatial discretization,
leading to discretized derivatives and, consequently, higher order
derivative corrections in the effective Hamiltonian.  The
characteristic size of the discretization $\epsilon$ in
\cite{QFTonCST} is (as $a$ here) a weighted mean
$\epsilon\approx\lP^\alpha\lambda^{1-\alpha}$. While $\alpha$ was
not uniquely fixed there, $\alpha=2/3$ would reproduce the result
here. Also, the order of the first correction due to the higher
derivative terms was, exactly as in our case, found to be
$\epsilon^2/\lambda^2$.

Next, since \cite{QFTonCST} also works with coherent
states, there is a parameter (called $a$ there) corresponding
closely to our parameter $d$, distributing the width of the state
between configuration and momentum degrees of freedom. There, this
parameter is chosen macroscopic. One can probably understand this
from our result here that the correction \eqref{fluctu} decreases
while the correction \eqref{lattice} increases with increasing
$d$. However, the decrease is governed by a higher power of $d$
such that larger $d$ are favored.

Finally, the relative order of the correction due to quantum effects
found in \cite{QFTonCST} is
$(\epsilon/\lambda)^{2/\alpha-1}$. Again, this corresponds precisely
to our $(\lP/a)^4$ for $\alpha=2/3$. As for comparison with the
dispersion relations in \cite{Correct} the analysis proceeds
similarly.

\section{The Issue of Lorentz Invariance}

A case in which modified dispersion relations have good chance of
being tested is when they break Lorentz invariance. This allows
correction terms of the form $\lP E$ which can be high enough for
sufficiently large energy $E$. Possible Lorentz invariant corrections,
on the other hand, can at most be of the order $\lP m$ with the fixed
and limited mass $m$. Accordingly, except for trivially modified
dispersion relations that have been discussed for quantum gravity
phenomenology, all break Lorentz invariance.

Loop quantum gravity in particular  is a Hamiltonian formalism
where Lorentz invariance is not manifest. (The supposedly
covariant twin of loop quantum gravity, spin foams, is under much
less control currently and not yet suitable for explicit
applications; moreover, anomaly free formulations may even loose
manifest covariance \cite{Anomaly}.)
If we first consider only the spatial aspects, rotational
invariance is not manifestly broken by the discrete structure
since one does not restrict the theory to a fixed lattice.
Nevertheless, calculations of dispersion relations choose a fixed
graph which cannot be rotationally invariant, but do the
calculations in such a way that the result is rotationally
symmetric. So a priori discreteness does not necessarily imply
violations of symmetries in the approximate classical expressions.
For Lorentz invariance, however, this is much more difficult to
achieve since time does not appear directly in the theory.

The Hamiltonian formulation requires calculations to be based on a
lattice in space such that only the spatial geometry is manifestly
discrete. Nevertheless, dispersion relations are computed from
classical field equations which involve coordinate time. This time
parameter is introduced by computing the perturbative matter
Hamiltonian on the lattice, and then treating it as the Hamiltonian of
a classical field theory. Time then appears via the Hamiltonian
equations of motion, but only at the classical level. In particular,
time is always continuous in this setting unlike space, whose discrete
structure is responsible for the very effects to be computed.

That the situation in loop quantum gravity is indeed such that the
calculations done so far introduce Lorentz violations not coming from
the theory is suggested by an additional complication for this kind of
question caused by discrete theories. Discrete theories are non-local
which implies that they have effective formulations of higher
derivative type (when differences are to be expanded in a Taylor
series to arrive at an effective Hamiltonian or action). For higher
derivative theories, in turn, the Legendre transform does not commute
with a perturbation expansion: If we start with a higher derivative
Lagrangian in which the higher derivative terms can be treated as
perturbations, and Legendre transform, then the resulting Hamiltonian
will not be analytic in the perturbation parameters \cite{Simon}. If,
on the other hand, we first Legendre transform the full Lagrangian
then the perturbation expansion of the resulting Hamiltonian must
obviously be analytic in the perturbation parameter.

While a Lagrangian formulation would immediately show whether
or not Lorentz invariance is broken by correction terms, the
Hamiltonian formulation is more indirect. Since perturbation and
Legendre transform do not commute, it is in general not viable to
compute corrections to the Hamiltonian and then Legendre transform
to find an effective Lagrangian to read off possible Lorentz
violations. The corrected Hamiltonian itself would not be of
higher order in time derivatives and so the corresponding
Lagrangian would be analytic in the expansion parameters. But if
higher derivatives for the full expressions have to be expected,
the Legendre transform of the perturbed Hamiltonian would not
coincide with the perturbed full Lagrangian. In particular if
there are higher spatial derivatives in the Hamiltonian, as in any
spatially discrete theory, there are two possibilities much more
complicated to distinguish perturbatively: Either there are no
higher time derivatives in the corresponding full Lagrangian,
which would break Lorentz symmetry since there are higher space
derivatives; or there are higher time derivatives, in which case
the theory may or may not be Lorentz invariant.

We illustrate these issues with an example for a discrete theory with
finitely many degrees of freedom. Let the action be $S=\sum_n\epsilon
L(q_n,q_{n+1})$ with Lagrangian
$L(q_n,q_{n+1})=(q_{n+1}-q_n)^2/2\epsilon^2$, a discretization of a
free particle with discrete time step $\epsilon$. If we define the
momentum by $p_n:=\epsilon\partial L/\partial q_{n+1}$ and the
Hamiltonian by $L(q_n,q_{n+1})=p_n(q_{n+1}-q_n)/\epsilon- H(q_n,p_n)$, we
obtain $p_n=(q_{n+1}-q_n)/\epsilon$ and $H(q_n,p_n)=p_n^2/2$.

We now assume that $\epsilon$ is small and approximate the discrete
values $q_n$ by a continuous function $q(t)$ such that $q_n=q(\epsilon
n)$. Thus,
\[
 L(q)=\frac{1}{2}
\left(\sum_{k=1}^{\infty}\frac{\epsilon^{k-1}}{k!}q^{(k)}\right)^2=
\frac{1}{2}(\dot{q}^2+\epsilon \dot{q}\ddot{q}+
\epsilon^2(\tfrac{1}{3}\dot{q} q^{(3)}+ \tfrac{1}{4}\ddot{q}^2)+
O(\epsilon^3))
\]
which yields a higher derivative theory at second order in $\epsilon$,
which we will use in what follows. (To linear order in $\epsilon$,
however, the theory is not higher derivative since the only correction
is a total derivative.) Only $q$ itself and the first derivative
$\dot{q}$ are independent variables since $q^{(3)}$ can be removed
from the Lagrangian by integrating by parts. Removing all the total
derivatives and higher orders in $\epsilon$ results in the Lagrangian
\[
 L(q)=\frac{1}{2}\dot{q}^2- \frac{1}{24}\epsilon^2 \ddot{q}^2
\]
which after performing a (higher derivative) Legendre transform gives momenta
\begin{eqnarray*}
 \pi_q &:=& \frac{\partial L}{\partial\dot{q}}-\frac{\md}{\md t}
\frac{\partial L}{\partial\ddot{q}}= \dot{q}+\frac{1}{12}\epsilon^2
q^{(3)}\\
 \pi_{\dot{q}} &:=& \frac{\partial L}{\partial\ddot{q}}=
-\frac{1}{12}\epsilon^2\ddot{q}
\end{eqnarray*}
and the Hamiltonian
\begin{equation} \label{Hdisc}
 H(q,\pi_q,\dot{q},\pi_{\dot{q}})=
\dot{q}\pi_q+\ddot{q}\pi_{\dot{q}}-L=
\dot{q}\pi_q-\frac{1}{2}\dot{q}^2-
6\epsilon^{-2}\pi_{\dot{q}}^2\,.
\end{equation}
As anticipated, the Hamiltonian is not analytic in $\epsilon$.

If we had started in a Hamiltonian formulation and gotten our discrete
formulation there, as in loop quantum gravity, we would have proceeded
differently. First, expanding a Hamiltonian formulation does not
introduce new degrees of freedom such as $\dot{q}$ above, which is
independent of $q$ and has its own momentum. Moreover, the symplectic
structure would be left untouched since it is independent of the
Hamiltonian (while a Lagrangian formulation mixes symplectic structure
and dynamics). Thus, we would still work with the unperturbed momentum
$\pi_q=\dot{q}$ for the only degree of freedom $q$. This fact can also
be derived from the Hamiltonian (\ref{Hdisc}) by solving the
Hamiltonian equation of motion $\dot{q}=\partial H/\partial\pi_q$ for
$\pi_q(\dot{q})$ which is now assumed not to be an independent
variable. We obtain
$\dot{q}=\dot{q}-(\dot{q}-\pi_q)\md\dot{q}/\md\pi_q$, which, since now
$\dot{q}$ is by assumption no longer independent of $\pi_q$,
immediately gives $\pi_q=\dot{q}$.

The second difference is that only perturbative corrections to the
Hamiltonian could appear, and not the $1/\epsilon$-term above. If we
remove this term and use the unperturbed momentum, we obtain
$H=\frac{1}{2}\pi_q^2$, which coincides with the continuous
approximation of the full discrete Hamiltonian. Thus, to this order of
perturbation theory, the Hamiltonian would not show any corrections
unlike the Lagrangian. The reason is that in this example the
perturbative corrections are all of higher derivative form, which
cannot be seen in the Hamiltonian.

The example can easily be extended in such a way that also the
Hamiltonian receives perturbative corrections. We now use two discrete
coordinates and Lagrangian $L=((q_{m,n+1}-q_{m,n})^2-
(q_{m+1,n}-q_{m,n})^2)/2\epsilon^2$ where we interpret $m$ as a
discrete space coordinate and $n$ as discrete time, as above. The
Lagrangian is symmetric under the exchange of $m$ with $n$, which
mimics a space-time symmetry. The momentum of $q_{m,n}$ now is
$p_{m,n}=(q_{m,n+1}-q_{m,n})/\epsilon$ and the Hamiltonian
$H=\frac{1}{2}(p_{m,n}^2+ (q_{m+1,n}-q_{m,n})^2/\epsilon^2)$.

Perturbing the Lagrangian leads to
$L=\frac{1}{2}(\dot{q}^2-(q')^2+
\epsilon(\dot{q}\ddot{q}-q'q'')+ \epsilon^2(\tfrac{1}{3}\dot{q}
q^{(3)}+ \tfrac{1}{4}\ddot{q}^2-\tfrac{1}{3}q' q'''-
\tfrac{1}{4}(q'')^2)+
O(\epsilon^3))$. The perturbed momenta are as before, and the Legendre
transform of the perturbed Lagrangian with total derivatives removed
as before yields
\[
 H(q,\pi_q,\dot{q},\pi_{\dot{q}})= \dot{q}\pi_q-\frac{1}{2}\dot{q}^2+
\frac{1}{2}(q')^2 + \frac{1}{24}\epsilon^2 (q'')^2 -
6\epsilon^{-2}\pi_{\dot{q}}^2
\,.
\]
With the prescription above, a perturbation of the Hamiltonian would
have led to the analytic expression
$H=\frac{1}{2}(\pi_q^2+(q')^2+\frac{1}{12}\epsilon^2 (q'')^2)$, which
only shows the higher order correction (which is of higher derivative
form only in space but not in time), but not the non-analytic
correction coming from the higher derivative nature.

These examples have important hints for the calculation of corrected
dispersion relations and the issue of Lorentz covariance. Since only
higher order corrections will be seen when a Hamiltonian is
perturbed, Lorentz violations are bound to appear as a consequence
of this way of doing the calculation. Space and time derivatives of
the classical fields have to be related in the Lagrangian in a way
dictated by the symmetry. If those terms are torn apart, because one
computes the Lagrangian from a perturbed Hamiltonian which only sees
higher space derivatives but not higher time derivatives in its
corrections, Lorentz invariance will be violated. This kind of
violation of Lorentz symmetry is not a consequence of the theory but
of the way to perform perturbative calculations.

We present one more example showing the role of higher derivatives in
Lorentz invariant theories. We use the Lagrangian
\[
 L=-\frac{1}{2}\int(\psi(\Box+\epsilon\Box^2)\psi+m^2\psi^2)
\]
for a scalar of mass $m$. It leads to the field equation
\[
 -(\Box+\epsilon\Box^2)\psi=m^2\psi
\]
which is Lorentz invariant. The dispersion relation can be computed
from the plane wave ansatz $\psi(x,t)=\exp(i(Et-kx))$ and takes the
form
\[
 \epsilon E^4+(1-2\epsilon k^2)E^2+k^2(\epsilon k^2-1)=0
\]
such that
\[
 E^2=k^2-\frac{1}{2\epsilon}\pm\sqrt{1+4\epsilon m^2}/2\epsilon\,.
\]
If $\epsilon\ll m^{-2}$, we obtain
\[
 E^2=k^2-\frac{1}{2\epsilon}(1\mp(1+2\epsilon m^2-2\epsilon^2m^4))
\]
with two non-analytic solutions, which have to be discarded in a
perturbative situation, and the corrected relation
\[
 E^2=k^2+m^2-\epsilon m^4+O(\epsilon^2)
\]
which is Lorentz invariant.

The Hamiltonian situation of this example is as follows. We have
momenta $\pi_{\psi}=
\dot{\psi}-2\epsilon\Delta\dot{\psi}+\epsilon\dddot{\psi}$ and
$\pi_{\dot{\psi}}=-\epsilon\ddot{\psi}$ leading to the Hamiltonian
\[
 H=-\frac{1}{2}\dot{\psi}^2+\dot{\psi}\pi_{\psi}-
\frac{1}{2}\psi\Delta\psi+ \frac{1}{2}m^2\psi^2+
\epsilon(\dot{\psi}\Delta\dot{\psi}+ \psi\Delta^2\psi/2)-
\frac{1}{2}\epsilon^{-1}\pi_{\dot{\psi}}^2\,.
\]
Again, this is non-analytic in $\epsilon$, but would lead to Lorentz
invariant equations of motion. If, on the other hand, we had started
with perturbing a Hamiltonian, we could only have seen the analytic
part and would not have introduced new degrees of freedom and instead
used $\pi_{\psi}=\dot{\psi}$ (this again also follows from the
equations of motion under the assumption of having no additional
degrees of freedom: $\dot{\psi}=\partial H/\partial\pi_{\psi}=
-\dot{\psi}\partial\dot{\psi}/\partial\pi_{\psi}+\dot{\psi}+
\pi_{\psi}\partial\dot{\psi}/\pi_{\psi}$ implies immediately
$\pi_{\psi}=\dot{\psi}$ if, as per our assumption,
$\partial\dot{\psi}/\partial\pi_{\psi}\not=0$). Thus, we would have
arrived at a Hamiltonian
\[
 H=\frac{1}{2}(\pi_{\psi}^2-\psi\Delta\psi+ m^2\psi^2+
\epsilon(2\pi_{\psi}\Delta\pi_{\psi}+ \psi\Delta^2\psi)
\]
and perturbed equation of motion
\[
 \ddot{\psi}=\Delta\psi-m^2\psi+\epsilon(\Delta^2\psi-2m^2\Delta\psi)\,.
\]
The dispersion relation for this equation is
\[
 E^2=k^2+m^2-\epsilon k^2(k^2+2m^2)
\]
which does break Lorentz invariance.

\section{Conclusions}

The observations presented here beg the question of what is the
correct procedure to compute modified dispersion relations from a
Hamiltonian point of view when higher derivative terms have to be
expected. Such a procedure has to be amended incorporating the
semiclassical dynamics in a more controlled way. This in particular
has to take care of new degrees of freedom that emerge from higher
derivative theories. One possibility is to derive a full,
non-perturbative discrete Hamiltonian from the quantum theory, which
is understood as a classical object but on a discrete space e.g. 
\cite{PullinGambini}. Before
one expands and computes equations of motion, one has to transform to
a Lagrangian, also on discrete space and time. From then on one can
work with perturbative expansions and compute modified dispersion
relations.

There are obvious difficulties in the way of implementing this
procedure in loop quantum gravity since already calculations with a
perturbed Hamiltonian are cumbersome. It should however be kept in
mind that the calculations done up to now (including the model of the
previous section) can only yield preliminary results and that a
definite answer to the question of Lorentz violation by loop quantum
gravity definitely has to await a more complete treatment, possibly
along the lines sketched above. Alternatively, perturbative
Hamiltonian techniques for effective actions, which also allow to see
additional degrees of freedom coming from higher derivatives, can be
developed. This approach, which is now under investigation, would
allow to perform the perturbation expansion at the Hamiltonian level
all the time.

We expect that the model presented in the first part of this paper
can be used for a first step in applying those methods to the issue of
dispersion relations. As we showed, it shares most qualitative and
even some quantitative features with more elaborate calculations and
thus is simple but reliable. It can therefore play a role in deriving
modified dispersion relations that better take into account the higher
derivative nature.

\section{Acknowledgments}

We are grateful to Aureliano Skirzewski for discussions and comments
on a draft. It is a pleasure to acknowledge the support and warm
hospitality during visits to the Perimeter Institute and The
Pennsylvania State University where this work was completed. HAMT has
been partially supported by CONACYT-40745-F; this work was supported
in part by the Eberly Research Funds of Penn State and by the NSF
grant PHY-00-90091.


\end{document}